\documentclass[11pt]{article}  
\usepackage{graphicx,floatflt,amssymb}    
\usepackage{epsfig}    
\usepackage{axodraw}

\def\dfrac#1#2{{\displaystyle {#1 \over #2}}}

\def\IJMPA #1 #2 #3 {{\sl Int.~J.~Mod.~Phys.}~{\bf A#1}\ (19#2) #3$\,$}
\def\MPLA #1 #2 #3 {{\sl Mod.~Phys.~Lett.}~{\bf A#1}\ (19#2) #3$\,$}
\def\NPB #1 #2 #3 {{\sl Nucl.~Phys.}~{\bf B#1}\ (19#2) #3$\,$}
\def\PLB #1 #2 #3 {{\sl Phys.~Lett.}~{\bf B#1}\ (19#2) #3$\,$}
\def\PR #1 #2 #3 {{\sl Phys.~Rep.}~{\bf#1}\ (19#2) #3$\,$}
\def\JHEP #1 #2 #3 {{\sl JHEP}~{\bf #1}~(19#2)~#3$\,$}
\def\PRD #1 #2 #3 {{\sl Phys.~Rev.}~{\bf D#1}\ (19#2) #3$\,$}
\def\PTP #1 #2 #3 {{\sl Prog.~Theor.~Phys.}~{\bf #1}\ (19#2) #3$\,$}
\def\PRL #1 #2 #3 {{\sl Phys.~Rev.~Lett.}~{\bf#1}\ (19#2) #3$\,$}
\def\RMP #1 #2 #3 {{\sl Rev.~Mod.~Phys.}~{\bf#1}\ (19#2) #3$\,$}
\def\ZPC #1 #2 #3 {{\sl Z.~Phys.}~{\bf C#1}\ (19#2) #3$\,$}
\def\PPNP#1 #2 #3 {{\sl Prog. Part. Nucl. Phys. }{\bf #1} (#2) #3$\,$}
\newcommand{\ba}{\begin{array}}     
\newcommand{\ea}{\end{array}}     
\newcommand{\bd}{\begin{displaymath}}     
\newcommand{\ed}{\end{displaymath}}     
\newcommand{\be}{\begin{equation}}     
\newcommand{\ee}{\end{equation}}     
\newcommand{\bea}{\begin{eqnarray}}     
\newcommand{\eea}{\end{eqnarray}}

\def\eVq{{\rm eV}^{2}}

\begin{document}     
\vspace*{-0.5in}     
\renewcommand{\thefootnote}{\fnsymbol{footnote}}     
\begin{flushright}     
LPT Orsay/07-97   
\end{flushright}     

\vskip 5pt     

\begin{center}     
{\Large {\bf Neutrinos and Lepton Flavour Violation in the \\ Left-Right Twin Higgs Model}}
\vskip 25pt     
{Asmaa Abada} and         
{Irene Hidalgo}       
\vskip 10pt      
{\small  
{\it Laboratoire de Physique Th\'eorique,  UMR 8627,     
Universit\'e de Paris-Sud XI, B\^atiment 210, \\
91405 Orsay Cedex, France} }

\normalsize     
\end{center}

\vskip 20pt     

\begin{abstract}  

We analyse the lepton sector of the Left-Right Twin Higgs Model. This model offers an alternative way to solve the ``little hierarchy" problem of the Standard Model. 
We show that one can achieve an effective see-saw to explain the origin of neutrino masses and that this model can accommodate the observed neutrino masses and mixings. We have also studied the lepton flavour violation process $\ell_1\to \ell_2\gamma$ and discussed how the experimental bound from these branching ratios constrains the scale of symmetry breaking of this Twin Higgs model. 

\vskip 5pt \noindent  
\texttt{Key Words: Neutrinos, Twin Higgs, Lepton Flavour Violation}  
\end{abstract}

\setcounter{footnote}{0}  
\renewcommand{\thefootnote}{\arabic{footnote}}  

\section{Introduction}
 The Higgs mass stabilisation is among the most important  motivations  to search for physics beyond the
Standard Model (SM). The naturalness problem associated with  the Higgs mass is known as the
SM ``hierarchy problem" \cite{hierarchyproblem}, and it is due to large quadratic radiative corrections to 
the Higgs mass. If the latter is ``naturally" of the order of the electroweak (EW) scale (i.e., not the result of an accidental cancellation between higher scales), then new physics that compensates these dangerous quadratic contributions should appear at the scale of a few TeV. Among the various candidates for  new physics based on  extensions of the Higgs sector, there is the recent proposal of  Twin Higgs models \cite{THmirror,THLR,phenotwin}. In the latter models, the Higgs mass is protected at the one-loop level by new symmetries and  another Higgs field, called the Twin Higgs, is introduced. The Twin Higgs mechanism proceeds in two main steps: i) the SM Higgs emerges as a pseudo-Goldstone boson from a spontaneously broken global symmetry, similar to what  happens in the Little Higgs models \cite{LH}; ii) an additional discrete symmetry is imposed, in such a way that the leading quadratically divergent terms cancel each other, and do not contribute to the Higgs mass. The resulting Higgs mass is naturally of order of the EW scale, with a cut-off for the theory around 10 TeV. 

There are two ways to  implement the Twin Higgs mechanism. Either the additional symmetry is a mirror parity (implying that a copy of the SM is introduced~\cite{THmirror}), or a Left-Right symmetry~\cite{THLR,phenotwin}. In the present work, and since we are interested in the study of neutrinos with both chiralities, we consider the second possibility.

In addition to the theoretical problems of the SM, there are experimental 
evidences that suggest the existence of new physics, among which is  the observation of 
neutrino oscillations \cite{oscillneutrinos}. The latter observation implies  
that neutrino have masses and mix. However, the extreme smallness  of  neutrino masses suggests that their origin may  differ from that of the other fermions. 
 The most natural explanation for 
this lightness is given by the see-saw mechanism, in which right-handed 
neutrinos are introduced with Majorana masses, $M$, larger than the 
electroweak scale.

In this work we  will explore the leptonic sector of this Left-Right Twin Higgs (LRTH) model, discussing  whether neutrino masses and mixings can be generated. In particular, we study   how a light neutrino spectrum, generated via an effective see-saw,  can be embedded in the LRTH  framework. 
 As we will see, this model offers a rich phenomenology, in particular there are additional sources of lepton flavour violation (LFV).  We will consider the $\mu \rightarrow e \gamma$ process, analysing the phenomenological implications on the Higgs sector arising  from the constraints associated with the experimental bound on the branching ratio.

This work is organised as follows: in Section 2 we present the relevant 
ingredients of the Left-Right Twin Higgs model. Section 3 is devoted to 
the generation of neutrino masses in the one-generation case, while Section 4 is dedicated to 
the three-generation case. In Section 5 we consider the LFV processes $\ell_1\to \ell_2\gamma$,  in particular 
the $\mu \rightarrow e \gamma$ decay. We finally
summarise our results in Section 6. 

\section{Left-Right Twin Higgs Model}

In the Left-Right Twin Higgs model, the global symmetry is U(4)$\times$U(4) which is spontaneously broken to U(3)$\times$U(3), and explicitly broken by the gauging of an SU(2)$_{L} \times$SU(2)$_{R} \times$U(1)$_{B-L}$ subgroup. The twin symmetry is identified with a ``Left-Right" parity which interchanges $L$ and $R$,  implying that gauge and Yukawa couplings of SU(2)$_{L}$ and SU(2)$_{R}$ are identical (e.g., $g_{2L} = g_{2R}$).

In the fundamental representation of each U(4), 
the Higgs field can be written as  $H$ = $( H_{L}, H_{R} )$ , and the Twin Higgs as  $\hat{H}$ = $( \hat{H}_{L},\hat{H}_{R} )$. The fields $H_{L}$  and $\hat{H}_{L}$ are  charged under SU(2)$_{L}$, while $H_{R}$  and $\hat{H}_{R}$ are  charged  under SU(2)$_{R}$. 
 
After both Higgses develop vacuum expectation values (vevs), 
\begin{equation}<H> = (0,0,0,f)\ ,\quad <\hat{H}> = (0,0,0,\hat{f})\, , \label{vevs}
\end{equation}
the global symmetry U(4)$\times$U(4) breaks to U(3)$\times$U(3), while SU(2)$_{R} \times$U(1)$_{B-L}$ breaks down to the SM U(1)$_{Y}$. We must further consider the electroweak breaking of SU(2)$_{L} \times$U(1)$_{Y}$. 
After the breaking scheme is finalised, six Goldstone bosons are eaten by the massive gauge bosons: the SM  $Z$ and $W$, and two extra heavier bosons, $Z_{H}$ and $W_{H}$. We are left with one neutral pseudoscalar, $\phi^0$, a pair of charged scalars $\phi^\pm$, the SM physical Higgs $h$, and an ${\rm SU}(2)_L$ Twin Higgs doublet $\hat{h}=(\hat{h}_1^+, \hat{h}_2^0)$.

We begin by describing the gauge sector. 
The gauge fields are $(W_L^\pm,W_L^0)$ for ${\rm SU}(2)_L$ and $(W_R^\pm,W_R^0)$ for ${\rm SU}(2)_R$, and there is a $W_1$  gauge field corresponding
to U(1)$_{B-L}$. After the successive symmetry breakings, there are six massive gauge bosons $W$, $W_H$, $Z$, $Z_H$, and one massless photon $\gamma$.  In  the charged gauge bosons,  there is
no mixing between the $W_L$ and the $W_R$: $W=W_L$
and $W_H=W_R$. The neutral gauge bosons $Z_H$, $Z$ and $\gamma$ are linear combinations of $W_L^0$, $W_R^0$ and $W_1$. At the tree-level, the masses of the heaviest gauge bosons are:
\begin{equation}
   m^2_{W_H} = \frac{1}{2}g_2^2 (\hat{f}^2+f^2) ,\ \ \
   m_{Z_H}^2 =\frac{g_1^2+g_2^2}{g_2^2}(m^2_{W_H}+m^2_{W})-m_{Z}^2\ ,
\end{equation} 
where $g_1$ and $g_2$ are the gauge couplings for ${\rm U}(1)_{B-L}$ and
${\rm SU}(2)_{L,R}$.

The fermion sector is similar to the SM, with the addition of three right-handed neutrinos. The quarks and leptons are charged under ${\rm SU}(3)_c\times {\rm
SU}(2)_L\times {\rm SU}(2)_R\times {\rm U}(1)_{B-L}$ as
\begin{eqnarray}
    L_{L}=-i\left(\begin{array}{c}~\nu_{L}
\\l_{L}\end{array}\right):
({\bf 1},{\bf 2},{\bf 1},-1), \ \ \ \ \ &&
    L_{R}=\left(\begin{array}{c}~\nu_{R}
\\l_{R}\end{array}\right): 
({\bf 1},{\bf 1},{\bf 2},-1),\nonumber\\
    Q_{L}=-i\left(\begin{array}{c} u_{L}
\\ d_{L}\end{array}\right): 
({\bf 3},{\bf 2},{\bf 1},1/3),\ \ \ \ \ &&
    Q_{R}=\left(\begin{array}{c}u_{R}
\\d_{R}\end{array}\right):({\bf 3},{\bf 1},{\bf 2},1/3)\ .
\end{eqnarray}

Notice from the above equation that  we now have doublets under ${\rm SU}(2)_R$.

Fermions acquire masses via non-renormalisable dimension 5 operators. For the quark sector, the non-renormalisable operators  have the form
\begin{equation}
    \frac{y_u}
{\Lambda}(\bar{Q}_{L}\tau_2 H_L^*)(H_R^T\tau_2{Q}_{R})
+\frac{y_d}{\Lambda}(\bar{Q}_{L}
H_L)(H_R^{\dagger}{Q}_{R})
+ {\rm {H.c}}.,
 \label{eq:Yukawa1}
\end{equation}
where $\tau_2=-i\sigma_2$, $\sigma_2$ being the Pauli matrix.
Once $H_R$ obtains a vev, these non-renormalisable couplings reduce to effective quark Yukawa couplings
of the order of $f/\Lambda$. This mechanism works for small Yukawa couplings, but not in the case of the ${\cal O}(1)$ 
top Yukawa coupling. Therefore, in order to solve this, a pair of vector-like quarks are introduced, with the following quantum numbers:
\begin{equation}
    T_L: ({\bf 3},{\bf 1},{\bf 1},4/3),\ \ \ \ \
    T_R: ({\bf 3},{\bf 1},{\bf 1},4/3).
\end{equation}

Then, we can write the gauge invariant interactions
\begin{equation}
    y_L\bar{Q}_{L3}\tau_2 H_L^*T_R+y_R\bar{Q}_{R3}\tau_2H_R^*T_L - M_T\bar{T}_LT_R + {\rm {H.c.}}\ .
\label{eq:topyukawa}
\end{equation}

Under the Left-Right symmetry, $y_L=y_R=y$.
Once the  $H_{L,R}$ Higgses get vevs,
the first two terms in Eq.~(\ref{eq:topyukawa}) generate masses
for an SM-like top quark $(u_{L3},q_R)$, with mass $yv/\sqrt{2}$ ($v$ being the SM-Higgs vev), and
a heavy top quark $(q_L, u_{R3})$ with mass $yf$. A non-zero value of the mass of the $T$ vector-like quark $M_T$ leads to the mixing between the SM-like top
quark and the heavy top quark. Provided $M_T \lesssim f$ and that $y$ is of order one, the  top Yukawa will also be of order one.

Similar interactions can be written for the other scalar field $\hat{H}$.  However, the heavy top quark will get a much larger
mass of the order of $y\hat{f}$, reintroducing the fine-tuning problem in the Higgs potential. To avoid this, a parity is imposed in
the model, under which $\hat{H}$ is odd, while all the other fields are
even. This ``$\hat{H}$-parity" thus forbids renormalisable couplings between
$\hat{H}$ and fermions, especially the top quark. Therefore,
at renormalisable level, 
$\hat{H}$ only couples to the gauge boson sector, while $H$
couples to both the gauge sector and the matter fields.

The phenomenology of the gauge and quark sector of the Left-Right Twin Higgs model has been studied in \cite{phenotwin}. As we are going to see in following sections, our work is focused on the phenomenology of the leptonic sector, which we proceed to describe.

\section{Neutrino mass generation in Left-Right Twin Higgs Model}

In addition to the theoretical problems of the SM, there is  experimental 
evidence that suggests the existence of new physics: the observation of 
neutrino oscillations \cite{oscillneutrinos}, implying  
that neutrinos are massive and mix. The observed smallness of the neutrino masses suggests that their origin may  differ from that of the other fermions. 
 
The lepton sector of the LRTH model contains 3 generations of  ${\rm SU}(2)_{L,R}$ doublets 
which are charged under ${\rm SU}(3)_c\times {\rm SU}(2)_L\times {\rm SU}(2)_R\times {\rm U}(1)_{B-L}$ as 
\begin{eqnarray}
    L_{L\alpha}=-i\left(\begin{array}{c}~\nu_{L\alpha}
\\l_{L\alpha}\end{array}\right):
({\bf 1},{\bf 2},{\bf 1},-1), \ \ \ \ \ &&
    L_{R\alpha}=\left(\begin{array}{c}~\nu_{R\alpha}
\\l_{R\alpha}\end{array}\right): 
({\bf 1},{\bf 1},{\bf 2},-1),\nonumber
\end{eqnarray}
where  $\alpha$, the family index, runs from 1 to 3. Due to the Left-Right symmetry, it is mandatory to introduce  three
generations of right-handed neutrinos $\nu_{R\alpha}$, which combine with $l_{R\alpha}$ to form ${\rm
SU}(2)_R$ doublets.

The charged leptons obtain their masses in the same way as the first two generations of quarks, i.e. via non-renormalisable dimension 5 operators, which for the lepton sector are

\begin{equation}
{y_l^{ij}\over \Lambda} (\bar{L}_{Li} H_L)(H_R^{\dagger}{L}_{Rj})+{y_{\nu}^{ij}\over \Lambda} (\bar{L}_{L,i}\tau_2 H_L^*)(H_R^T\tau_2{L}_{Rj}) + {\rm {H.c.}} ,
\label{eq:Yukawalep}
\end{equation}
where $i$ and $j$ run from 1 to 3.  Analogous to what occurs in the quark sector, after the Higgs $H_R$ develops a vev, one obtains effective Yukawa couplings  $y_{\nu,l}^{ij} f/\Lambda$, which will give rise to  Dirac mass terms once $H_L$ acquires a vev. Still, this is not enough to explain the extreme smallness of the neutrino masses ($m_{\nu} \leq$ 1 eV).

The left- and right-handed neutrinos can be of Majorana nature and the related  Majorana mass terms can also be generated via dimension 5 operators. For simplicity, and considering only the one generation case, these operators induce the following terms
\bea
{c_{L}\over \Lambda}\, \left( \overline{L}_{L} \tau_2 H_{L}^\dagger \right)^2+ {\rm H.c} \qquad, \qquad {c_{R} \over \Lambda}\, \left( \overline{L}_{R} \tau_2 H_{R}^\dagger \right)^2+ {\rm H.c.}~.
\label{MajoranaLR}\   
\eea

Due to the Left-Right parity, the couplings for the right and left sectors are  the same $c_L=c_R=c$. Notice that these operators 
 will induce Majorana mass terms for both neutrino chiralities. Thus, it is natural to envisage the possibility
of a see-saw like mechanism~\cite{seesaw} to explain the smallness of the light neutrino masses.

With the terms from Eqs. (\ref{eq:Yukawalep}) and (\ref{MajoranaLR}) one can easily   see that it is hardly  possible to achieve an effective see-saw inducing  very small masses for the left-handed neutrinos and heavy masses for the right-handed ones.
Therefore, we need a new ingredient, and this comes from the twin Higgs $\hat{H}$. The $\hat{H}$-parity that forbids $\hat{H}$ to couple to the fermions  can be broken for  the neutrino sector. Indeed,  this parity was only introduced in order to prevent the heavy top quark from getting a large mass of order $y \hat{f}$ (which would consequently reintroduce the fine-tuning in the Higgs mass), and 
it is not mandatory that the $\hat{H}$-parity still holds in the lepton sector. More specifically, we will assume that the twin Higgs $\hat{H}_R$ couples to the right-handed neutrinos.
Thus, it is possible to have the following term:
\bea
{c_{\hat{H}} \over \Lambda}\, \left( \overline{L}_{R} \tau_2 \hat{H}_{R}^\dagger \right)^2+ {\rm H.c.}~,
\label{dimfiv}  
\eea 
which will give a contribution to the Majorana mass of the right-handed neutrino, in addition to those of  Eq.~(\ref{MajoranaLR}).

Once $H_R$ and $\hat{H}_R$ get vevs, respectively $f$ and $\hat{f}$ (Eq.~(\ref{vevs})), and after EW symmetry breaking, we can derive the following see-saw mass matrix\footnote{Since only $\hat{H}_R$ from the twin sector gets a vev and does couple to neutrinos, we do not have an additional Majorana contribution to the ($\nu_L$,$\nu_L$)  entry in ${\cal M}$.} for the LRTH model in the basis ($\nu_L$,$\nu_R$): 
\bea
{\cal M} = \left(\begin{array}{cc}
c {v^ 2 \over 2\Lambda}  & y_{\nu} {v f \over \sqrt{2}\Lambda} \\
y_{\nu}^{T} {v f \over \sqrt{2}\Lambda} & c {f^ 2 \over \Lambda}+ c_{\hat{H}} {\hat{f}^ 2 \over \Lambda}
\end{array}\right)~.
\label{calM}
\eea

%%%%%%%%%%%%%%%%%figure%%%%%%%%%%%%%%%%%%%%%%%%
\begin{figure}[t]
\vspace{0.2cm}
\hspace{0.5cm}
\centerline{\psfig{figure=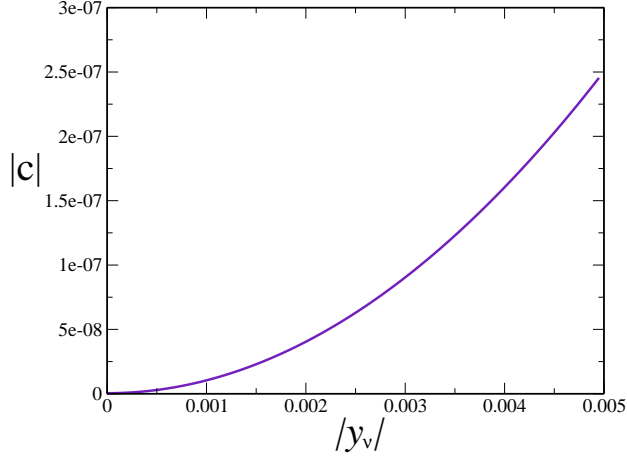,height=9.5cm,angle=-90}
 }
\caption{\footnotesize ($|c|$, $|y_{\nu}|$) parameter space (see  Eq.~(\ref{nulight}) )  fulfilling  the constraint  $m_{\nu_{light}}\lesssim 1$ eV.}
\label{couplings1gen}
\end{figure}
%%%%%%%%%%%%%%%%%%%%%%%%figure%%%%%%%%%%%%%%%%%%%%%%%%

In the one-generation case one has two massive states, a heavy ($\sim \nu_R$) and a light one. Taking into account that $v <  f < \hat{f}$, the masses of the eigenstates are:

\be
\label{nuheavy} 
m_{\nu_{heavy}} = c {f^ 2 \over \Lambda}+c_{\hat{H}} {\hat{f}^ 2 \over \Lambda} + {\cal O}\left({f^ 2 v^ 2 \over \hat{f}^ 2 \Lambda}\right) \simeq c_{\hat{H}} {\hat{f}^ 2 \over \Lambda} + {\cal O}\left({f^ 2 v^ 2 \over \hat{f}^ 2 \Lambda}\right),
\ee

\be
\label{nulight}
m_{\nu_{light}} = {v^ 2 \over 2\Lambda} \left( c-y_{\nu}^{T} c_{\hat{H}}^{-1} y_{\nu} { f^ 2 \over \hat{f}^ 2} \right) + {\cal O}\left({f^ 2 v^ 4 \over \hat{f}^ 4 \Lambda}\right).
\ee

Let us consider a typical choice for the parameters:  $\Lambda = 4 \pi f$, $f \sim$ 1 TeV and $c_{\hat{H}}$ of ${\cal O}(1)$. Then, once $f$ is fixed, $\hat{f}$ is determined by the minimisation of the Higgs potential \cite{THLR,phenotwin}, i.e. $\hat{f} \sim 10 f$, leading to  $m_{\nu_{heavy}} \sim$ 10 TeV. The experimental bounds on the mass scale of the light neutrinos is around 1 eV. Imposing this bound, we plot in Fig.~\ref{couplings1gen} the variation in the modulus of $c$ versus $y_{\nu}$ of Eq. (\ref{nulight}). If we want to avoid fine-tuning\footnote{The absence of fine-tuning in this case occurs when the two  terms with opposite sign in Eq.~(\ref{nulight}) are of the same order, or smaller, than the light neutrino mass, $m_{\nu_{light}}$. The most conservative way to avoid fine-tuning is to take an absolute mass for $m_{\nu_{light}}$ that saturates the cosmological bound~\cite{cosmo}, i.e. $m_{\nu_{light}}\sim$ 0.3 eV.} between the two terms in Eq.~(\ref{nulight}), then the Yukawa couplings for the neutrino, $y_{\nu}$, must be of ${\cal O}(10^{-3})$ (the same order as the charged lepton Yukawa couplings), but $c$ must be much smaller, of ${\cal O}(10^{-7})$.
 This is manifest in Fig.~\ref{couplings1gen}. However, nothing forbids the coefficient  $c$ to be set to zero, which corresponds to the case where lepton number violation only takes place  in the twin sector. Should this happen, we are led to a standard see-saw (of type-I \cite{seesaw}), with heavy neutrinos of the order of 10 TeV.

\section{Study of the three generations case}

In the previous section we have considered the simple one-generation case. We now generalise our study to the three-generation case, in order to see whether or not the LRTH model can accommodate  the present neutrino data~\cite{oscillneutrinos,cosmo}. The general couplings for three generations are obtained  by promoting  the couplings  $y_{\nu}$,  $c$ and $c_{\hat{H}}$ to 3 $\times$ 3 matrices in flavour space, ${\bf Y}_{\nu}$, ${\bf C}$ and ${\bf C}_{\hat{H}}$.

In the basis where the  charged lepton mass matrix is diagonal, the 3 $\times$ 3 neutrino mass matrices are:

\be
\label{nuheavy3x3} 
{\bf M}_{\nu_{heavy}} \simeq {\bf C}_{\hat{H}} {\hat{f}^ 2 \over \Lambda} ,
\ee

\be
\label{nulight3x3}
{\bf M}_{\nu_{light}} \simeq {v^ 2 \over 2\Lambda} \left( {\bf C}-{\bf Y}_{\nu}^{T} {\bf C}_{\hat{H}}^{-1} {\bf Y}_{\nu} { f^ 2 \over \hat{f}^ 2} \right) .
\ee

One can always take ${\bf M}_{\nu_{heavy}}$, i.e. $ {\bf C}_{\hat{H}}$,  to be diagonal in this basis,  $ {\bf C}_{\hat{H}}={\rm diag}(c_1,c_2,c_3)$.
On the other hand, the diagonalisation of ${\bf M}_{\nu_{light}}$  is given by the PMNS~\cite{UPMNS} matrix, $U_{PMNS}$, which in the standard parametrisation is

\be
\label{3_mixing}
U_{PMNS} = 
\left(\matrix{ 
c_{13}c_{12} & c_{13}s_{12} & s_{13}e^{-i\delta} \cr 
-c_{23}s_{12} -s_{23}s_{13}c_{12}e^{i\delta} & 
c_{23}c_{12} -s_{23}s_{13}s_{12}e^{i\delta} & s_{23}c_{13} \cr 
s_{23}s_{12} -c_{23}s_{13}c_{12}e^{i\delta} & 
-s_{23}c_{12} -c_{23}s_{13}s_{12}e^{i\delta} & c_{23}c_{13} \cr}
\right)\; \times {\rm diag} \left(1, e^{i\alpha}, e^{i\beta}\right)\,,
\ee
where $ s_{ij}\equiv \sin \theta_{ij}$, $ c_{ij}\equiv \cos \theta_{ij}$, and $\delta$, $\alpha$ and $ \beta$ are CP violating phases. After the diagonalisation of ${\bf M}_{\nu_{light}}$, one has three light Majorana neutrino mass eigenstates $\nu_i$, with small masses $m_{\nu_i}$.

From neutrino oscillation experiments \cite{oscillneutrinos}, the allowed ranges for the  neutrino observable  parameters at $1\sigma$ level ($3\sigma$ level) are:
\bea
\label{eq:3nuranges}
    \;\Delta m^2_{21}\;
    &=& 7.9\,_{-0.28}^{+0.27}\,\left(_{-0.89}^{+1.1}\right)
    \times 10^{-5}~\eVq \,,
    \\
    \left|\Delta m^2_{31}\right|
    &=& 2.6 \pm 0.2\,(0.6) \times 10^{-3}~\eVq \,,
    \\
    \theta_{12} &=& 33.7\pm 1.3\,\left(_{-3.5}^{+4.3}\right) \,,
    \\
    \theta_{23} &=& 43.3\,_{-3.8}^{+4.3}\,\left(_{-8.8}^{+9.8}\right) \,,
    \\
    \theta_{13} &=& 0\,_{-0.0}^{+5.2}\,\left(_{-0.0}^{+11.5}\right) \,,
\eea
and the cosmological upper bound~\cite{cosmo} on the sum of neutrino masses is $\sum_i |m_{\nu_i}| <$ 1 eV. Given the observed frequencies, $\Delta m^2_{ij}$, there are three possible patterns for the mass eigenvalues:

\bea {\tt{Degenerate}}& : & |m_1|\sim |m_2| \sim |m_3|\gg |m_i-m_j|\nonumber\\ {\tt{Normal~hierarchy}}& : & |m_1|\sim
|m_2| \ll |m_3| \nonumber\\ {\tt{Inverted~hierarchy} }& : & |m_1|\sim
|m_2| \gg |m_3|
\label{abc}
\eea  
 
The experimental data will constrain the coupling matrices  ${\bf Y}_{\nu}$, ${\bf C}$ and ${\bf C}_{\hat{H}}$. In order to discuss this, we organise the analysis as  follows. First, we consider two cases of spectrum in the right-handed sector: degenerate heavy neutrinos and hierarchical heavy neutrinos. In both cases, we choose typical textures for the coupling matrices and see how the latter  are constrained by experiment. In all cases, we ignore the CP phases, as our main goal is to constrain the magnitude of the Yukawa couplings.

\begin{itemize}

\item Degenerate heavy neutrinos.

\vspace{0.1cm}
\begin{table}[!t]
\begin{center}
%\resizebox{13cm}{2.7cm} {
\begin{tabular}{|c|c|c|c|}   
\hline  
&{\tt Degenerate} & {\tt Normal} & {\tt Inverted}\\
\hline
\phantom{\huge I}$|\lambda|$ & [0-4.5]$\times 10^{-5}$&  [0-4.8]$\times 10^{-5}$& 5$\times 10^{-6}$-3.5$\times 10^{-5}$\\
\hline
\phantom{\huge I}$|\delta|$ & [0-3]$\times 10^{-5}$&  5$\times 10^{-6}$-2$\times 10^{-5}$& 5$\times 10^{-6}$-4$\times 10^{-5}$\\
\hline
\phantom{\huge I}$|\eta|$ & 1.3$\times 10^{-12}$-1$\times 10^{-11}$& & \\
\hline
\phantom{\huge I}$|\alpha|$ & 6.5$\times 10^{-12}$-2$\times 10^{-11}$& $10^{-11}$& 7.5$\times 10^{-12}$-4.5$\times 10^{-11}$\\
\hline
\phantom{\huge I}$|\beta|$ & 6.5$\times 10^{-12}$-2$\times 10^{-11}$&  8$\times 10^{-12}$-3 $\times 10^{-11}$& 1$\times 10^{-11}$-3$\times 10^{-11}$\\
\hline
\phantom{\huge I}$|\epsilon|$ & 5$\times 10^{-12}$-3$\times  10^{-11}$&  [0-7.5]$\times 10^{-12}$& 1.5$\times 10^{-11}$-4$\times 10^{-11}$\\
\hline
\end{tabular}
%} 
\end{center}
\caption{Derived ranges for the ${\bf Y}_\nu$ and ${\bf C}$ texture parameters in  the case of degenerate heavy neutrinos. The light neutrino spectrum can verify all  possible patterns.}
\label{tab01}
\end{table}

This case corresponds to ${\bf C}_{\hat{H}}$ equal to the identity matrix,  leading to degenerate masses for the heavy neutrinos of the order of 10 TeV. For example, we can use the triangular parametrisation for the Yukawa matrix  ${\bf Y}_{\nu}$~\cite{Branco}. A simple choice to reduce the number of parameters and that still provides a representative overview of the solutions, is

\be
{\bf Y}_\nu =
\left(
\begin{array}{ccc} \lambda & 0 & 0\\ \delta & \lambda& 0\\ \delta & \delta & \lambda
\end{array}
\right)~~~.
\label{Ytriangular}
\ee 

Then, the distinction between the different patterns is given by the parametrisation of the ${\bf C}$ matrix. We will consider the following typical textures~\cite{Altarelli} for  ${\bf C}$:

\be 
{\bf C}_{degenerate} =
\left(
\begin{array}{ccc} \eta & \epsilon & \epsilon \\ \epsilon & \eta +\alpha& \epsilon\\ \epsilon & \epsilon & \eta +\beta
\end{array}
\right)~~~,
\label{Cdegen}
\ee

\be 
{\bf C}_{normal}=
\left(
\begin{array}{ccc} \alpha & \epsilon & \epsilon \\ \epsilon & \beta& \beta\\ \epsilon & \beta & \beta
\end{array}
\right)~~~,
\label{Cnormal}
\ee 

\be 
{\bf C}_{inverted} =
\left(
\begin{array}{ccc} \alpha & -\epsilon & \epsilon \\ -\epsilon & \beta& \beta\\ \epsilon & \beta & \beta
\end{array}
\right)~~~.
\label{Cinverted}
\ee
 
We derive the allowed ranges for the modulus of the parameters of these 
matrices from the requirement of compatibility with experimental data.  The  obtained results are shown in Table~1.

\item Hierarchical heavy neutrinos.

\vspace{0.1cm}
\begin{table}[!t]
\begin{center}
%\resizebox{8.5cm}{2.8cm} {
\begin{tabular}{|c|c|c|}   
\hline   
& {\tt Normal} & {\tt Inverted}\\
\hline
\phantom{\huge I}$|\lambda|$ & [0-3]$\times 10^{-7}$&  [0-5.3]$\times 10^{-7}$\\
\hline
\phantom{\huge I}$|\delta|$ & [0-1]$\times 10^{-6}$&  [0-1]$\times 10^{-6}$\\
\hline
\phantom{\huge I}$|\eta|$ & 9$\times 10^{-13}$-6$\times 10^{-12}$& [0-1]$\times 10^{-12}$ \\
\hline
\phantom{\huge I}$|\alpha|$ & 1$\times 10^{-11}$-2$\times 10^{-11}$& [0-3]$\times 10^{-11}$\\
\hline
\phantom{\huge I}$|\beta|$ & 1$\times 10^{-11}$-2.5$\times 10^{-11}$&  5$\times 10^{-12}$-3.5$\times 10^{-11}$\\
\hline
\phantom{\huge I}$|\epsilon|$ & 8.5$\times 10^{-12}$-1$\times 10^{-11}$&  2$\times 10^{-11}$-4.7$\times 10^{-11}$\\
\hline
\end{tabular}
%}
\end{center}
\caption{Derived ranges for the ${\bf Y}_\nu$ and ${\bf C}$ texture parameters in  the case of hierarchical heavy neutrinos. The light neutrino spectrum cannot be degenerate in this case.}
\label{tab02}
\end{table}

In this case, the choice of the hierarchy is arbitrary, so we have assumed 
the simple choice, $c_1:c_2:c_3 = 10^{-4}:10^{-2}:1$. With the Yukawa matrix  ${\bf Y}_{\nu}$ of Eq.~(\ref{Ytriangular}), we have found that only the ${\bf C}$ matrix parametrisation of Eq.~(\ref{Cdegen}) is allowed by experimental constraints. This leads only to a hierarchical light neutrino spectrum, and a degenerate spectrum cannot be obtained. The corresponding allowed ranges for the parameters are  given in Table~2.

\end{itemize}

As we have mentioned in the one-generation case,  if we assume  that the only  source of  lepton number violation, $\slash$ $\hspace{-.37cm}{\rm L}$, lies in the twin sector, the coupling $c$ must be set to zero in Eq.~(\ref{nulight}). In the three-generation case, this corresponds to a coupling matrix ${\bf C}$ equal to 0. From now on, we perform the analysis under this  assumption.  We therefore use for  ${\bf Y}_{\nu}$  the general triangular parametrisation~\cite{Branco}  given by

\be 
{\bf Y}_\nu =
\left(
\begin{array}{ccc} \alpha & 0 & 0\\ \beta & \kappa& 0\\ \delta & \epsilon & \lambda
\end{array}
\right)~~~,
\label{Ytriangeneral}
\ee 

and we proceed as above, considering  two cases of  spectra in the right-handed sector: degenerate and hierarchical. In Table 3  we show the derived ranges for the parameters of the triangular parametrisation when the experimental constraints are considered, for both normal and inverted hierarchy for light neutrinos. We have found  that there is no possible way to achieve a degenerate spectrum for the light neutrinos without an important fine-tuning.

\vspace{0.1cm}
\begin{table}[!t]
\begin{center}
%\resizebox{13.5cm}{3.5cm} {
\begin{tabular}{|c|c|c||c|c|}   
\hline  
& \multicolumn{2}{|c||}{Degenerate Heavy $\nu$} &   \multicolumn{2}{|c|}{Hierarchical Heavy $\nu$} \\
\hline
& {\tt Normal} & {\tt Inverted}&  {\tt Normal} & {\tt Inverted}   \\
\hline
\phantom{\huge I}$|\alpha|$ & [0-2.5]$\times 10^{-5}$&[1.2-1.8]$\times 10^{-5}$& [0-1.2]$\times 10^{-7}$&[0-4]$\times 10^{-7}$\\
\hline
\phantom{\huge I}$|\beta|$ & [0-1.2]$\times 10^{-5}$&3$\times 10^{-5}$& 9$\times 10^{-7}$-1.2$\times 10^{-6}$&[1.4-3.4]$\times 10^{-6}$\\
\hline
\phantom{\huge I}$|\kappa|$ & [2-3]$\times 10^{-5}$&3.5$\times 10^{-5}$& 2$\times 10^{-6}$ & [3-4]$\times 10^{-6}$\\
\hline
\phantom{\huge I}$|\delta|$ & [0-1]$\times 10^{-5}$ & [1-5.6]$\times 10^{-6}$& [0-5]$\times 10^{-6}$ & 4$\times 10^{-7}$-3.2$\times 10^{-6}$\\
\hline
\phantom{\huge I}$|\epsilon|$ & [2-2.5]$\times 10^{-5}$ & 6$\times 10^{-7}$-6$\times 10^{-6}$& 2.3$\times 10^{-5}$ & [0-5.6]$\times 10^{-6}$\\
\hline
\phantom{\huge I}$|\lambda|$ & [3.6-4.2]$\times 10^{-5}$ & 4.6$\times 10^{-5}$ & [3.4-4]$\times 10^{-5}$ & [4.3-4.8]$\times 10^{-5}$\\
\hline
\end{tabular}
%}
\end{center}
\caption{On the left (right) side, derived ranges for the $Y_\nu$ texture parameters in  the case of degenerate (hierarchical) heavy neutrinos. The light neutrino spectrum cannot be degenerate in this  case  where lepton number violation occurs only in the twin sector.}
\label{tab03}
\end{table}

\section{Lepton flavour violation processes: $\mu \rightarrow e \gamma$}

The Lagrangian in Eq.~(\ref{eq:Yukawalep}) induces masses and mixings for neutrinos, and it may also be a source of lepton flavour violation. The Standard Model, even when minimally extended by right-handed neutrinos to accommodate neutrino masses,  predicts extremely small  branching ratios for charged LFV processes, namely BR($\mu \rightarrow e \gamma$) $< 10^{-50}$ \cite{SMLFV}.  The current experimental upper bound on the  process  $\mu \rightarrow e \gamma$ \cite{expLFV} is
 \be
 {\rm{BR}}(\mu \rightarrow e \gamma) < 1.2 \times 10^{-11}.
\label{expmuegamma}
\ee

In the LRTH model the flavour structure is richer than in the minimally extended SM. There are right-handed neutrinos that couple to Higgses and to  heavy gauge bosons, and this leads to an enhancement of the branching ratio for $\mu \rightarrow e \gamma$. In this section, we will study  the phenomenological consequences of this enhancement regarding  the scale of this framework.

In addition to the minimally extended SM with right-handed neutrinos diagram contributing to $\mu \rightarrow e \gamma$, diagram a) of Fig.~2, we have to consider  the contributions of the heavy gauge boson, $W_{H}$, and the charged Higgses, $\phi^\pm$ (respectively b), c) and d) of Fig.~2). The relevant vertex interactions for these processes are explicited in Fig.~3.

\begin{figure}[h]
\centering
\begin{picture}(330,150)(0,0)
\ArrowLine(10,80)(50,80)
\ArrowLine(50,80)(90,80)
\ArrowLine(90,80)(130,80)
\PhotonArc(70,80)(20,0,180){2}{10.5}
\Photon(70,100)(70,130){1.5}{5}
\Text(10,81)[bl]{$\mu$}
\Text(130,82)[br]{$e$}
\Text(74,130)[tl]{$\gamma$}
\Text(57,95)[br]{\small{$W^-$}}
\Text(88,95)[bl]{\small{$W^-$}}
\Text(70,82)[b]{\scriptsize{$\nu_i$}}
\Text(72,63)[br]{a)}
\ArrowLine(180,80)(220,80)
\ArrowLine(220,80)(260,80)
\ArrowLine(260,80)(300,80)
\PhotonArc(240,80)(20,0,180){2}{10.5}
\Photon(240,100)(240,130){1.5}{5}
\Text(180,81)[bl]{$\mu$}
\Text(300,82)[br]{$e$}
\Text(244,130)[tl]{$\gamma$}
\Text(226,95)[br]{\small{$W_H^-$}}
\Text(258,95)[bl]{\small{$W_H^-$}}
\Text(240,83)[b]{\scriptsize{$\nu_{Hi}$}}
\Text(242,63)[br]{b)}
\ArrowLine(10,0)(50,0)
\ArrowLine(50,0)(90,0)
\ArrowLine(90,0)(130,0)
\DashCArc(70,0)(20,0,180){3}
\Photon(70,20)(70,50){1.5}{5}
\Text(10,1)[bl]{$\mu$}
\Text(130,2)[br]{$e$}
\Text(74,50)[tl]{$\gamma$}
\Text(57,15)[br]{\small{$\phi^-$}}
\Text(88,15)[bl]{\small{$\phi^-$}}
\Text(70,2)[b]{\scriptsize{$\nu_i$}}
\Text(72,-18)[br]{c)}
\ArrowLine(180,0)(220,0)
\ArrowLine(220,0)(260,0)
\ArrowLine(260,0)(300,0)
\DashCArc(240,0)(20,0,180){3}
\Photon(240,22)(240,50){1.5}{5}
\Text(180,1)[bl]{$\mu$}
\Text(300,2)[br]{$e$}
\Text(244,50)[tl]{$\gamma$}
\Text(227,15)[br]{\small{$\phi^-$}}
\Text(258,15)[bl]{\small{$\phi^-$}}
\Text(240,3)[b]{\scriptsize{$\nu_{Hi}$}}
\Text(242,-18)[br]{d)}
\end{picture}
\vspace{.5cm}
\caption{One-loop diagrams for $\mu \rightarrow e \gamma$  in the LRTH model.}
\label{Fig:diagrams}
\end{figure}
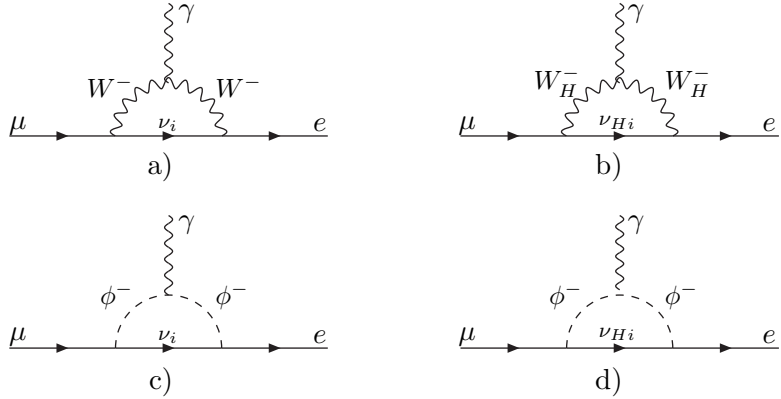

\begin{figure}[h]
\centering
\begin{picture}(330,170)(0,0)
\Photon(10,120)(70,120){1.5}{7}
\ArrowLine(70,120)(130,150)
\ArrowLine(130,90)(70,120)
\Text(40,122)[b]{\small{$W_{L,R}^-$}}
\Text(100,135)[rb]{$l_{L,R}^-$}
\Text(100,105)[rt]{$\nu_{L,R}$}
\Text(240,120)[]{$=\ \frac{e}{\sqrt{2} s_w}\gamma_{\mu}   P_{L,R}$}
\DashLine(10,30)(70,30){3}
\ArrowLine(70,30)(130,60)
\ArrowLine(130,0)(70,30)
\Text(40,32)[b]{\small{$\phi^-$}}
\Text(100,45)[rb]{$l_{L,R}^-$}
\Text(100,15)[rt]{$\nu_{L,R}$}
\Text(240,30)[]{$=\ \frac{i}{f}(m_{l_L,\nu_R} P_L-m_{\nu_{L},l_R}P_R)$}
\end{picture}
\caption{Vertex interactions for LFV processes.}
\label{Fig:interactions}
\end{figure}
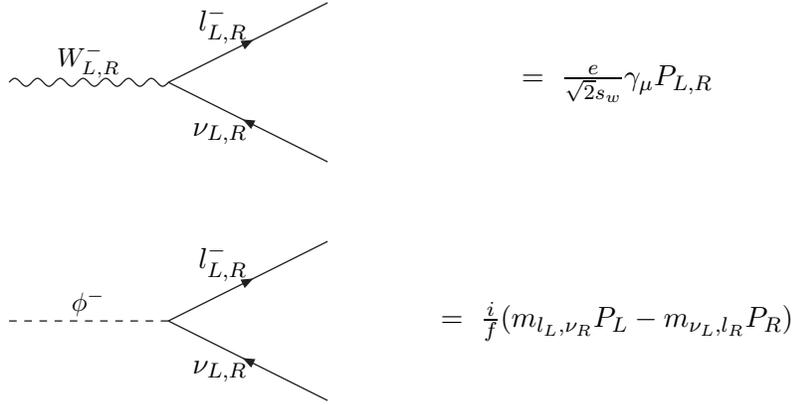

The amplitude for the $l_i\rightarrow l_j  \gamma$ process presents the
general form:
\bea
\label{T}
T=e \ \epsilon^\alpha\  \bar{l_j}\  {m_l}_i\  i \ \sigma_{\alpha \beta}\  q^\beta
\left(A_L P_L + A_R P_R \right)l_i\;, \eea
where $q$ and $\epsilon^\alpha$ are the momentum and the polarisation of the photon, $ P_{R,L}=(1\pm \gamma_5)/2$
and $A_L$ ($A_R$) (carrying $i,j$ indices) is the coefficient of the  amplitude for left (right) 
incoming $l_i$ lepton and thus  
right (left) $l_j$.  The
corresponding branching ratio is given by
\be
\label{BRgen}
{\mathrm{BR}}(l_i\rightarrow l_j  \gamma) =
\frac{12\pi^2 e^2}{G_F^2} \left(|A_L|^2 + |A_R|^2 \right) .
\ee

The diagrams contributing to $A_L$ (mediated by the light neutrinos, see Fig.~\ref{Fig:diagrams}, a) and c)) are clearly subdominant when compared to those mediated by the heavy  neutrinos  (see Fig.~\ref{Fig:diagrams}, b) and d)), which contribute to $A_R$. Thus, $A_L$ can be neglected and the resulting branching ratio for the process $\mu \rightarrow e \gamma$ can be written as
\be
\label{BR_AR}
{\mathrm{BR}}(\mu \rightarrow e  \gamma) =
\frac{12\pi^2 e^2}{G_F^2}  |A_R^{W_H}+A_R^{\phi^\pm}|^2 ,
\ee
where the amplitudes $A_R$ of Eq.~(\ref{BR_AR}) are given by the following expressions,
\bea
A_R^{W_H}&=&\frac{1}{32 \pi^2}\frac{g^2}{m^2_{W_H}}\sum_k V^{ke}_H V^{*k\mu}_H D\left( \frac{{m^{k^2}_{\nu_H}}}{m^2_{W_H}} \right) ,\\
A_R^{\phi^\pm}&=&\frac{1}{16 \pi^2}\frac{1}{f^2 m_{\phi^\pm}^2}\sum_k V^{ke}_H V^{*k\mu}_H {m^{k^2}_{\nu_H}} E\left( \frac{{m^{k^2}_{\nu_H}}}{m^2_{\phi^\pm}} \right) ,
\eea
with
\bea
m^2_{\phi^\pm}&=& \frac{3}{16\pi^2} g_1^2m_{W_H}^2(\ln\frac{\Lambda^2}{m_{Z_H}^2}+1),\\
D(x)&=&-\dfrac{3x^3}{2(x-1)^4}\log x +
\dfrac{2x^3+5x^2-x}{4(x-1)^3} ,\\
E(x)&=&\dfrac{x^2-1-2x\log x}{2(x-1)^3} ,
\eea
and the matrix $V_H$ parametrises the interactions of the  charged leptons with the heavy neutrinos, mediated by $W_H$ and $\phi^\pm$. In other words  $V_H$ is the leptonic mixing matrix for the right-handed sector\footnote{Recall that $W_H$ is equal to $W_R$.}. In the previous section, we have chosen the basis where the mass matrix for the heavy neutrinos is diagonal in flavour-space. If one assumes no mixing  in the right-handed  sector, $V_H$ is the identity matrix, leading to $A_R=0$. Since $A_L$ is extremely small, the associated BRs are orders of magnitude below the experimental bounds and no new constraints arise from the induced LFV processes in this case.  However, this is only an assumption and it could happen that there is mixing in the right-handed sector. 
To have an idea of the amount of LFV that one can get in this model and its phenomenological implications, we consider two representative  scenarios,  both of them  associated with a hierarchical  heavy neutrino spectrum  ($c_1:c_2:c_3 = 10^{-4}:10^{-2}:1$). These two scenarios differ in  the mixing matrix of the heavy sector, $V_H$, and once this matrix is fixed, the experimental bound in BR($\mu \rightarrow e \gamma$), Eq.~(\ref{expmuegamma}), implies a lower bound for the scale $f$. We will thus consider  the following two scenarios, A and B.

\begin{itemize}
\item[A)] In this scenario we assume  $V_H =V_{CKM}$ (mixing matrix of the quark sector \cite{CKM}, also in terms of the standard parametrisation), so that
\begin{equation}
\sin\theta_{12}=0.225 ,\qquad \sin\theta_{13}=0.0044 ,\qquad \sin\theta_{23}=0.042 ,\qquad \delta=65^\circ\,,
\end{equation}
and then depending on whether
\begin{itemize}
\item[i)] $\Lambda = 4 \pi f$  
\item[ii)] $\Lambda = 2 \pi f$ 
\end{itemize}
and considering the experimental bound of Eq.~(\ref{expmuegamma}), we have correspondingly obtained \begin{itemize}
\item[i)]  $f \gtrsim 0.6 {\rm \ TeV}$ 
\item[ii)] $f \gtrsim 3 {\rm \ TeV}$.
\end{itemize}

\item[B)] We assume in this case $V_H =U_{PMNS}$. In particular, we set the  parameters to
\be
\sin\theta_{12}=\sqrt{0.300}\,,\qquad \sin\theta_{13}=\sqrt{0.030}\,,\qquad \sin\theta_{23}=\frac{1}{\sqrt{2}}\,,\qquad \delta=65^\circ\,,
\ee
where the angles are consistent with the experimental data \cite{oscillneutrinos}, while the CP phase, $\delta$ is chosen  equal to the CKM phase (setting the phases $\alpha$ and $\beta$ to zero).
 In this case we have obtained: 
\begin{itemize}
\item[i)] $\Lambda = 4 \pi f\ \Rightarrow \ f \gtrsim 5 {\rm \ TeV}$
\item[ii)] $\Lambda = 2 \pi f\ \Rightarrow \ f \gtrsim 6 {\rm\  TeV}$.
\end{itemize}
 
\end{itemize}

The above found lower bounds for $f$  are in fact of the same order as the upper bounds obtained  from avoiding  fine-tuning in the EW symmetry breaking \cite{THLR,phenotwin}. Larger values of $f$ imply a larger fine-tuning (e.g. $f \sim 2$ TeV implies $\sim 5\%$ of fine-tuning  and $f\sim 1$ TeV implies $\sim 10\%$ of fine-tuning). In this sense, scenario~A) is favoured  with respect to B).

Together with the theoretical argumentations regarding the absence of fine-tuning, 
the phenomenological constraints from the leptonic sector (light neutrino spectrum and  experimental bounds on LFV BRs),  imply that, although considerably constrained,  there is still a viability window for the scale $f$, making the LRTH framework possible. For instance, allowing for a $5\%$ fine-tuning, $f \sim[ 0.6 \ -\  2]$~TeV.  
Recall that  we have chosen illustrative examples of mixings in the right-handed sector. 
However, in order to complete the study one may perform a general scan over the parameters of the model. The window  may then  be enlarged, but will be ultimately constrained from below by the Higgs mass bound.

\section{Summary and conclusions}

In this paper we have analysed the lepton sector of the Twin Higgs model with a  Left-Right symmetry (LRTH). This model was proposed as an alternative way to solve the ``little hierarchy" problem of the Standard Model. The study of the heavy gauge and quark sectors has previously  been done in detail in \cite{phenotwin}, suggesting a rich   collider phenomenology. 

The Dirac mass terms  for leptons and quarks have the same origin and are obtained via dimension 5 non-renormalisable operators. In addition to  the Dirac mass terms, Majorana mass terms for neutrinos can  also be achieved via dimension 5 operators. Thus,  heavy and light neutrinos acquire Majorana masses through a see-saw mechanism. In order to generate small neutrino masses, we have allowed the twin Higgs, $\hat{H}$, to couple to neutrinos by breaking the $\hat{H}$-parity. Since only $\hat{H}_R$ gets a vev, its coupling to neutrinos gives an additional  Majorana mass term for the right-handed neutrinos, resulting in an effective see-saw. 

We have  studied the three-generation case, trying  to accommodate neutrino  data for the two kinds of right-handed neutrino spectra: degenerate and  hierarchical. We have performed the analysis with different typical textures for the Yukawa matrices, and derived the  allowed ranges for the parameters  from experimental constraints.

The last part of the work was dedicated to the study of lepton flavour violating processes, in particular to the $\mu \rightarrow e \gamma$ process. Due to the presence of heavy gauge bosons and charged Higgses, there are new interactions that contribute to enhance the  branching ratio. Taking into account the experimental bound on this branching ratio and assuming new sources of mixing in the heavy sector, we have found a lower bound on the scale $f$ of the order of a few TeV, i.e. the same order as the upper bound given by considering the absence of fine-tuning in the Higgs mass. 
This allows us to severely constrain the scale of the LRTH model.

\section*{Acknowledgments}
We specially thank Stephane Lavignac for enlightening discussions and useful suggestions. We also
acknowledge Emi Kou and Ana Teixeira for useful comments on this work. 
The authors acknowledge the support of the Agence Nationale de la Recherche ANR through the project JC05-43009-NEUPAC.

\end{document}